# Breakdown of the Schmid law in bcc molybdenum related to the effect of shear stress perpendicular to the slip direction

R. Gröger[1,a], V. Vitek[1,b]

[1]University of Pennsylvania, Department of Materials Science and Engineering
3231 Walnut St, Philadelphia, PA 19104, USA

[a]groger@seas.upenn.edu, [b]vitek@lrsm.upenn.edu

**Abstract.** The breakdown of the Schmid law in bcc metals has been known for a long time. The asymmetry of shearing in the slip direction $\langle 111 \rangle$ in the positive and negative sense, respectively, commonly identified with the twinning-antitwinning asymmetry, is undoubtedly one of the reasons. However, effect of stress components other than the shear stress in the slip direction may be important. In this paper we investigate by atomistic modeling the effect of shear stresses perpendicular to the Burgers vector on the glide of $a/2\langle 111 \rangle$ screw dislocations. We show that these shear stresses can significantly elevate or reduce the critical resolved shear stress (*CRSS*) in the direction of the Burgers vector needed for the dislocation motion, i.e. the Peierls stress. This occurs owing to the changes of the core induced by these stresses. This effect may be the reason why slip systems with smaller Schmid factors may be preferred over that with the largest Schmid factor.

**Introduction**

Plastic behavior of body-centered cubic (bcc) metals, in particular transition metals of VB and VIB group, is known to be very different from that of metals with close-packed structures. This was first noted already in 1926 in G. I. Taylor's studies of the deformation of iron [1]. The distinguishing features are the rapid increase of both the yield and flow stress with decreasing temperature, tendency to cleavage at low temperatures, large strain-rate sensitivity, strong influence of interstitial impurities and, most remarkably, a tension-compression asymmetry detected in uniaxial loading tests of single crystals. When first observed, this asymmetry was attributed to the so-called twinning-antitwinning asymmetry of shearing in the slip direction $\langle 111 \rangle$ along $\{112\}$ planes [2]. However, many experimental investigations made in the last forty years suggest that also other stress components than the shear stress in the direction of slip may affect the tension-compression asymmetry.

Plastic flow of all bcc metals is characterized by the low mobility of $a/2\langle 111 \rangle$ screw dislocations. The reason, now well established, is that their core is not planar but spreads spatially which leads to a high lattice friction (Peierls) stress. Atomic level calculations have revealed that this spreading is principally into $\{110\}$ planes of the $\langle 111 \rangle$ zone; for reviews see [3-5]. However, the non-planar dislocation core also gives rise to other aspects of the plastic behavior of bcc crystals, most remarkably to the above-mentioned asymmetries. When a bcc crystal is mechanically loaded, there are two types of stress components that affect the glide of a particular dislocation: (a) Stress components that exert the Peach-Koehler force on the dislocation and can induce the dislocation glide, and (b) stress components that do not exert any Peach-Koehler force but can still alter the structure of the core. The latter stress components, called *non-glide stresses*, may influence significantly, via induced core changes, how large stresses exerting the Peach-Koehler force need to be applied for the dislocation to glide. In other words, the induced structural changes of the core modify the Peierls stress.

In order to elucidate this phenomenon we carried out computer modeling of the response of an $a/2[111]$ screw dislocation to applied stresses. This calculation was performed as follows. Periodic boundary conditions were imposed parallel to the dislocation line while perpendicular to the dislocation the block of atoms consisted of an active region, in which all the atoms were fully relaxed, and an inactive region where the atoms were permanently displaced in accordance with the anisotropic elastic field of the dislocation. When a stress is applied, strain field corresponding to this stress is linearly superimposed. The active region contained 720 atoms and the inactive region 860 atoms. The interaction between the atoms was described by the recently developed bond-order potential for molybdenum that reflects the mixed metallic and covalent character of bonding in this transition metal [6].

The initial configuration always corresponded to a fully relaxed single screw dislocation. In every calculation we first applied a shear stress perpendicular to the Burgers vector that may alter the core. Then, with this non-glide shear stress imposed, we gradually increased the shear stress parallel to the Burgers vector until the dislocation started to glide. The latter stress was applied as the pure shear stress in a *maximum resolved shear stress plane* (MRSSP) and its value at which the glide commenced was identified with the *critical resolved shear stress* (*CRSS*) for slip. This procedure allows us to determine the dependence of the *CRSS* on the shear stress perpendicular to the Burgers vector. Moreover, these calculations have been carried out for several orientations of the MRSSP so that the concomitant effect of the glide and non-glide shear stresses has been studied in detail. The results demonstrate that it is the combined effect of the asymmetry of shearing in the slip direction, which can be identified with the twinning-antitwinning asymmetry, and core changes induced by the stresses perpendicular to the Burgers vector, that is responsible for the complex slip behavior of single crystals of bcc molybdenum. The most conspicuous is the breakdown of the Schmid law and possibility of preference for slip on the systems with lower Schmid factors over that with the highest Schmid factor.

**Variation of the *CRSS* with the shear stress perpendicular to the slip direction**

The stress tensor applied to the block of atoms with the $a/2[111]$ screw dislocation in the middle was composed of the shear stress perpendicular and the shear stress parallel to the slip direction [111]. For the shear stress parallel to the Burgers vector, the orientation of the MRSSP is determined by the angle $\chi$ which this plane makes with the "reference" $(\bar{1}01)$ plane that was found to be the usual slip plane. Owing to the crystal symmetry, it is sufficient to consider $-30° \leq \chi \leq +30°$ which spans the angular region from the $(\bar{1}\bar{1}2)$ plane at $\chi = -30°$ to the $(\bar{2}11)$ plane at $\chi = +30°$.

When applying the stress we use a right-handed coordinate system, where the axis 3 is parallel to the direction of the dislocation line (and the Burgers vector), axis 2 is the normal to the MRSSP and axis 1 is perpendicular to both axes 2 and 3. The applied stress tensor is in this coordinate system

$$\Sigma = \begin{bmatrix} -\tau & 0 & 0 \\ 0 & \tau & \sigma \\ 0 & \sigma & 0 \end{bmatrix}, \tag{1}$$

where $\sigma$ is the shear stress in the MRSSP and the direction of the Burgers vector and $\tau$ determines the non-glide shear stress acting perpendicular to the slip direction.

Starting with $\sigma = 0$, the chosen value of $\tau$ was always built up incrementally with steps not larger than $0.005\,C_{44}$ to ensure a proper convergence of the atomic relaxations. When a given stress $\tau$ was attained, the block was subsequently subjected to the shear stress $\sigma$. Again, this loading was applied incrementally but with stress steps not larger than $0.001\,C_{44}$ since the changes in the dislocation core induced by $\sigma$ were significantly larger than those induced by $\tau$. The *CRSS* was then identified with the value of $\sigma$ at which the dislocation started to move through the block. At lower levels of $\sigma$ the dislocation core changes gradually from the sessile to the glissile configuration. When $\sigma = CRSS$, this transformation is complete and the dislocation moves.

**Results and discussion**

The calculations of the dependence of the *CRSS* on $\tau$ have been carried out for five different orientations of the MRSSP: $\chi = 0°$ (corresponding to the $(\bar{1}01)$ plane), $\chi = \pm 9°$ (corresponding to the $(\bar{6}15)$ and $(\bar{5}\bar{1}6)$ planes, respectively) and $\chi = \pm 19°$ (corresponding to the $(\bar{3}12)$ and $(\bar{2}\bar{1}3)$ planes, respectively). Because the results for $\chi = \pm 9°$ are qualitatively similar to those obtained for $\chi = \pm 19°$, we only present here the results for $\chi = 0°$ and $\chi = \pm 19°$. The $CRSS - \tau$ dependence is shown in Fig. 1 for the three orientations of the MRSSP.

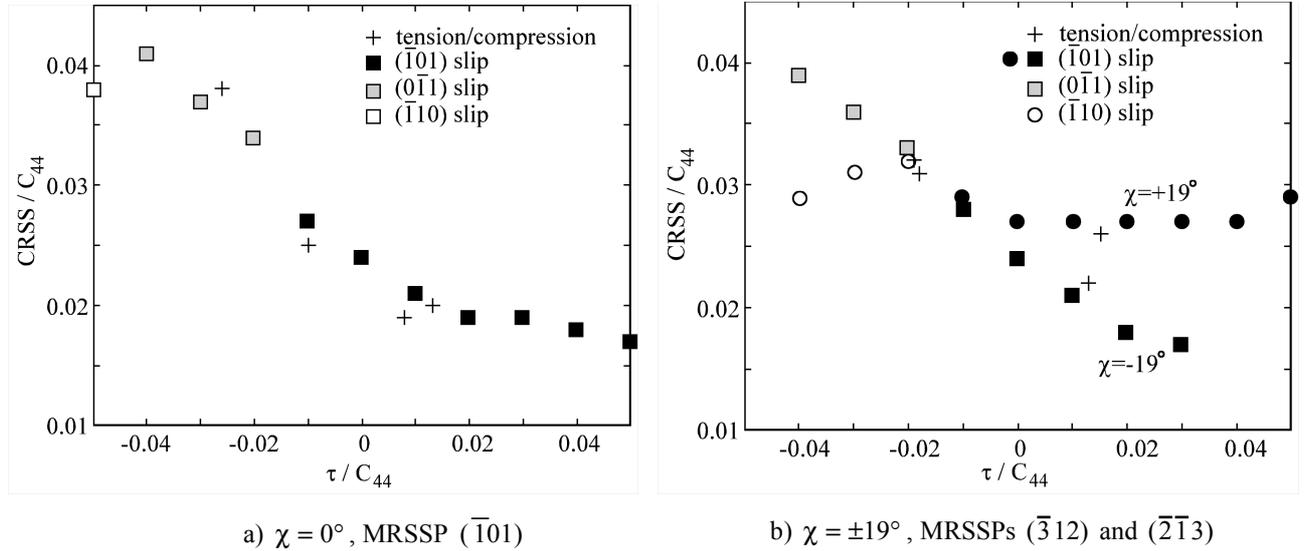

a) $\chi = 0°$, MRSSP $(\bar{1}01)$  b) $\chi = \pm 19°$, MRSSPs $(\bar{3}12)$ and $(\bar{2}\bar{1}3)$

Figure 1: Calculated dependence of the *CRSS* on $\tau$.

In general, the *CRSS* increases with decreasing $\tau$. Approximately, for $\tau/CRSS > -0.4$ the glide always occurred along the $(\bar{1}01)$ plane that possesses the highest Schmid factor in between the $\{1\bar{1}0\}$ planes of the [111] zone. On the other hand, for $\tau/CRSS < -0.4$ the slip was preferred on another $\{1\bar{1}0\}$ plane of the [111] zone, specifically either on the $(0\bar{1}1)$ or on the $(\bar{1}10)$ plane. The reason for this change of the slip plane can be understood by following the transformations of the dislocation core structure induced by application of $\tau$. In the core region this gives rise to significant displacements perpendicular to the Burgers vector. These displacements, equivalent to local edge components of the Burgers vector, then enhance synergistically the overall effect of $\tau$ on the core structure. Positive $\tau$ extends the dislocation core spreading on the $(\bar{1}01)$ plane and promotes thus the slip on this plane. In contrast, negative $\tau$ constricts the core spreading on the $(\bar{1}01)$ plane and extends it on both $(0\bar{1}1)$ and $(\bar{1}10)$ planes. Consequently, at high negative $\tau$ the slip is preferred on these planes. However, as discussed below, in real situation the slip direction will also change and a different slip system will operate for these values of $\tau$.

For comparison, we also show in Fig. 1 (crosses) the resolved shear stresses resulting from loading in tension/compression for several orientations of the loading axis within the stereographic triangle $[001]-[011]-[\bar{1}11]$. It is evident that the pairs $(\tau, CRSS)$ corresponding to these uniaxial loadings, correlate closely with the $(\tau, CRSS)$ obtained for the combination of $\tau$ and $\sigma$ when applied independently. These findings strongly support the idea that it is the shear stress perpendicular to the slip direction that modifies the *CRSS* for glide of an $a/2\langle 111 \rangle$ screw dislocation for loadings of any kind. The orientation dependencies of the *CRSS*, and related breakdown of the Schmid law, are then results of the combined effect of the asymmetry of the shear parallel to the Burgers vector and influence of shear stresses perpendicular to the Burgers vector.

Unlike in the atomistic model where only the dislocation with the $a/2[111]$ Burgers vector is present, in real crystals there are always dislocations with all four possible $a/2\langle 111 \rangle$ Burgers vectors. Hence, in reality, when applying the stress tensor $\boldsymbol{\Sigma}$, given by Eq. 1, glide of dislocations with Burgers vectors other than $a/2[111]$ may be induced. For example, for $\chi = 0°$ (Fig. 1a) the

*CRSS* for the glide on the $(\bar{1}01)$ plane is attained for $\sigma/C_{44} = 0.019$ when $\tau/C_{44} = +0.03$. At the same time, this combined loading induces for the slip direction $[\bar{1}11]$ shear stresses $\tau/C_{44} = +0.015$ and $\sigma/C_{44} = 0.028$ with the MRSSP inclined by $\chi = -20.6°$ with respect to the (110) plane. Since the slip system $(110)[\bar{1}11]$ is crystallographically equivalent to the system $(\bar{1}01)[111]$, we find from Fig. 1b (since there is no significant difference between $\chi = -19°$ and $\chi = -20.6°$) that for $\tau/C_{44} = +0.015$ the *CRSS* is attained for $\sigma/C_{44} = 0.018$. Thus, $a/2[\bar{1}11]$ dislocations will start to glide before $a/2[111]$ dislocations even though the Schmid stress for the $(\bar{1}01)[111]$ slip system is higher. Such transition to different slip systems will occur even more readily for larger negative and positive $\tau$ and thus the change of the slip plane for the [111] slip direction, observed for large negative $\tau$, will in reality correspond to occurrence of different slip systems.

**Conclusion**

We have shown by atomistic modeling that at $T = 0$ K the glide of $a/2\langle111\rangle$ screw dislocations in molybdenum, and undoubtedly other bcc metals, is controlled by both the shear stresses acting parallel and perpendicular to the slip direction. Presumably, at non-zero temperature the activation enthalpy associated with the dislocation motion will also depend on both these shear stresses. When the magnitude of shear stresses perpendicular to the Burgers vector is smaller than about $0.02C_{44}$ the {110}[111] slip system with the largest Schmid factor will operate though the Schmid law does not apply owing to the 'twinning-antitwinning' asymmetry of shear along the $\langle111\rangle$ direction. However, for larger shear stresses perpendicular to the Burgers vector another type of the breakdown of the Schmid law may occur: slip systems with smaller Schmid factors may be preferred.

The atomistic calculations presented here not only reveal the distinct non-glide shear stresses that affect the glide of $a/2\langle111\rangle$ screw dislocations, but the calculated *CRSS* - $\tau$ dependencies form a basis for multislip yield criteria and flow relations for continuum analyses which allow for the effect of non-glide stresses that originate at the level of individual dislocations to percolate up to the description of yielding in polycrystals [7, 8].

**Acknowledgements**

This research was supported by the National Science Foundation Grant no. DMR02-19243. Many discussions and comments of J. L. Bassani, V. Racherla and L. Yin are gratefully acknowledged.